\pdfoutput=1
\documentclass[sigconf, nonacm]{acmart}
\usepackage{enumitem}
\usepackage{amsmath}
\usepackage{tabularx}

\def\code#1{\texttt{#1}}

\AtBeginDocument{%
  }

\newcommand{\system}{NoteFlow}
\begin{document}

\title{\system{}: Leveraging Charts as Sight Glasses for Consistent and Continuous Data Flow Tracing}

\author{Yuan Tian}
\authornote{All authors are affiliated with the State Key Lab of CAD\&CG, Zhejiang University. }
\orcid{0009-0005-6089-7694}
\affiliation{
  \institution{Zhejiang University}
  \city{Hangzhou}
  \country{China}
}
\email{yuantian@zju.edu.cn}

\author{Dazhen Deng}
\authornote{Dazhen Deng is the corresponding author.}
\orcid{0000-0002-9057-8353}
\affiliation{
  \institution{Zhejiang University}
  \city{Hangzhou}
  \country{China}
}
\email{dengdazhen@zju.edu.cn}

\author{Sen Yang}
\orcid{0009-0000-9871-2644}
\affiliation{
  \institution{Zhejiang University}
  \city{Hangzhou}
  \country{China}
}
\email{seenyang@outlook.com}

\author{Huawei Zheng}
\orcid{0009-0004-0881-4292}
\affiliation{
  \institution{Zhejiang University}
  \city{Hangzhou}
  \country{China}
}
\email{zhenghuawei@zju.edu.cn}

\author{Bowen Shi}
\orcid{0009-0006-6089-380X}
\affiliation{
  \institution{Zhejiang University}
  \city{Hangzhou}
  \country{China}
}
\email{bwshi@zju.edu.cn}

\author{Kai Xiong}
\orcid{0000-0002-8203-9667}
\affiliation{
  \institution{Zhejiang University}
  \city{Hangzhou}
  \country{China}
}
\email{kaixiong@zju.edu.cn}

\author{Xinjing Yi}
\orcid{0009-0002-1874-1176}
\affiliation{
  \institution{Zhejiang University}
  \city{Hangzhou}
  \country{China}
}
\email{yixinjing@zju.edu.cn}

\author{Yingcai Wu}
\orcid{0000-0002-1119-3237}
\affiliation{
  \institution{Zhejiang University}
  \city{Hangzhou}
  \country{China}
}
\email{ycwu@zju.edu.cn}

\renewcommand{\shortauthors}{Tian et al.}

\begin{abstract}
Computational notebooks offer a flexible environment for exploratory data analysis (EDA), but this flexibility often leads to disorganized and iterative execution of notebook cells, making it difficult to track how data states evolve. Consequently, data scientists must devote extra mental effort to staying aware of data states, which is both tedious and prone to overlooking anomalies. To address this challenge, we developed NoteFlow, a notebook extension that leverages charts as ``sight glasses'' to provide a consistent and continuous tracing of data flow. NoteFlow allows users to (1) validate various facets of the current data state using recommended charts provided immediately after each cell execution, and (2) trace the global evolution of selected charts to continuously observe how particular data attributes evolve throughout the EDA process. We evaluated NoteFlow's effectiveness through a controlled study with 12 participants and a one-month field study with 2 data scientists on real-world workflows.

\end{abstract}


\begin{teaserfigure}
  \includegraphics[width=\textwidth]{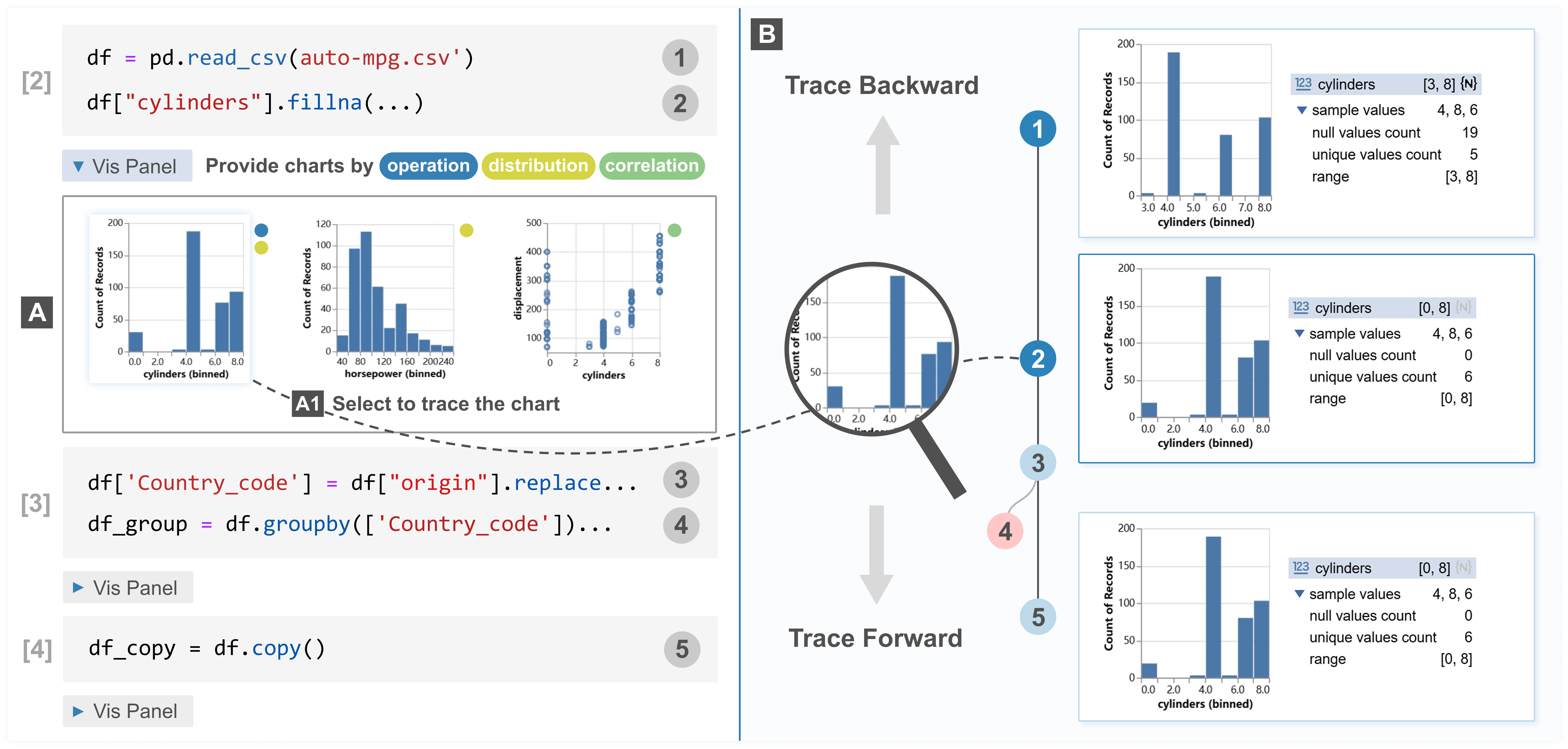}
  \caption{\system{} allows users to inspect provided charts after each cell (A) to understand the localized data state, and to freely select any chart (A1) as a ``sight glass'' to trace how a specific facet of the data evolves across the notebook (B).}
  \Description{The figure shows how NoteFlow uses charts to support data flow tracing in a notebook. After executing each cell, the system provides charts to summarize the localized data state. Users can then select any chart as a ``sight glass'' and trace the chart backward and forward across notebook cells to observe how that specific data facet evolves across the notebook.}
  \label{fig:teaser}
\end{teaserfigure}

\maketitle

\section{Introduction}
Exploratory Data Analysis (EDA) is important for data workers to uncover patterns, test hypotheses, and ensure data quality before formal modeling~\cite{wongsuphasawat2019goalsEDA, behrens1997principlesEDA}. 
EDA is inherently iterative and decision-driven, involving repeated cycles of data transformation, inspection of intermediate data tables, and decision-making for further exploration. 
This process gives rise to a dynamic \textbf{data flow}, composed of evolving data tables and their transformation lineage.

Computational notebooks, such as Jupyter Notebooks and R Markdown, are the most widely used tools for EDA~\cite{tukey1977exploratory}.
Their interactive and flexible design allows data workers to execute code incrementally, providing immediate feedback and enabling rapid implementation of new decisions. 
These features make computational notebooks particularly well-suited for the iterative and exploratory nature of EDA.
Despite their flexibility, computational notebooks provide limited built-in support for inspecting intermediate data tables. 
As a result, users often resort to writing custom code to visualize data, which can be tedious and error-prone. 
To ease this burden, several tools~\cite{lee2021lux, b2, epperson2022leveraging, autoprofiler, xavier} have introduced automated charting or profiling techniques to assist with the quick inspection of the latest data state.

However, EDA workflows typically involve a sequence of operations across cells, generating multiple intermediate tables with numerous attributes.
Even with automation, thoroughly inspecting each table remains cognitively demanding and time-consuming, 
increasing the risk of unnoticed issues propagating downstream and leading to anomalies in later analysis~\cite{2021datacascades, yangSubtleBugsEverywhere2021}.
This challenge stems from a fundamental limitation of current tools: they offer only localized, static snapshots that typically show the latest data state or immediate changes, without providing a cohesive view of how data evolves throughout the notebook.
As a result, when anomalies arise downstream, users are forced to scroll through previous cells, re-run code, and mentally reconstruct earlier states, which disrupts workflow and increases cognitive overhead.
The inherently flexible but non-linear execution of notebooks further complicates this process, making it difficult to trace the evolution of data across transformations and decisions.

To address these limitations, we explore \textbf{how charts can serve as consistent and continuous ``sight glasses''}, providing visual anchors that allow users to monitor, revisit, and reason about data evolution throughout the notebook.
We implement this idea in \system{}, a notebook extension designed for consistent and continuous data flow tracing, where data flow is defined as evolving data tables (represented as Pandas dataframes) and their transformation lineage.
Unlike prior approaches that focus on the local inspection of data, \system{} introduces a linked interface that connects notebook code with a global view of data flow. 
Charts are anchored to the evolving data flow and can be traced across transformations, enabling users to reason about
how the same analytical aspect changes over time rather than repeatedly inspecting disconnected snapshots.
Specifically, \system{} enables users to:
(1) inspect intermediate dataframe states through recommended charts, 
and (2) trace selected charts to understand how dataframes evolve throughout the analysis.
As the notebook progresses, traced charts are automatically updated, and their visual encodings are adapted to preserve semantic consistency across steps, enabling meaningful comparison.
With these features, \system{} supports both localized inspection and global understanding of data flow efficiently and cohesively.
Implementing such an extension introduces two key challenges:

\textbf{Scalable tracing of the entire EDA flow.}
In computational notebooks, a single transformation cell can generate multiple intermediate tables, making it difficult to trace all resulting states.
Flexible and non-linear cell execution further leads to tangled table relationships, complicating their visualization in a coherent layout.
Therefore, a core challenge is to scaleably visualize the explosion of intermediate tables and their relationships while maintaining global clarity. 
This calls for selectively de-emphasizing less relevant tables while keeping tables of interest in focus.
To this end, \system{} employs code parsing to extract and index intermediate tables along with their transformation lineage, and visualizes them in a node-link diagram where nodes represent tables and edges encode transformation relationships. 
The diagram adopts a top-down linear layout aligned with cell execution order, and supports selective table tracing while automatically collapsing irrelevant nodes, enabling scalable and structured data flow tracing.

\textbf{Effective slicing of the complex data space.} 
Charts offer a window into the data by visualizing specific attributes to reveal underlying patterns.
However, real-world tables often have many attributes, making exhaustive exploration impractical. 
While chart recommendation techniques~\cite{saket2018heuristics} help narrow this space, most existing methods~\cite{lee2021lux, epperson2022leveraging} focus on a single table and provide limited support for reasoning across the many intermediate tables created during EDA. 
Therefore, a key challenge is to maintain semantic consistency in chart recommendations across evolving tables, ensuring that visualizations remain comparable and interpretable as data transforms throughout the analysis. 
To address this, \system{} not only recommends charts for individual tables, but also organizes user-selected charts into a comparable view spanning multiple intermediate tables.
Specifically, \system{} preserves consistent semantic encodings for the selected charts across transformations, keeping the same visual mapping as the data changes.
This design constrains the chart space, makes data patterns comparable across tables, and enables users to effectively trace how data changes throughout the analysis.

We evaluated \system{} through a comparative user study and a one-month field study.
The comparative study shows that \system{} effectively supports tracing evolving data flow and maintaining awareness of both global and local data states. 
These benefits stem from its consistent chart encodings, dynamic tracing mechanisms, and the global overview of the data transformation process.
User feedback indicates that \system{}'s intuitive interface lowers the barrier to monitor data states, while its flexible tracing features support both backward and forward exploration.
The field study further complemented these findings by examining how \system{} was adopted in real-world data and scenarios, showing that local inspection reduced the overhead of data checks, chart tracing helped build clear mental models of data evolution, and these features collectively reduced unnecessary re-execution. 
It also revealed boundaries in \system{}'s current support.
We summarize key insights from both studies and discuss lessons learned to inform future work.
The contribution of this study includes:
\begin{itemize}[leftmargin=*]
\item We propose a conceptual framework for data flow tracing in computational notebooks that integrates flow parsing, chart recommendation, and chart-based tracing. 
\item We design an interface that links data flow with notebook code, with charts rendered over the flow to support tracing data evolution across the analysis. We implement this design in \system{}\footnote{\url{https://github.com/ZJUIDG-AIVA/NoteFlow}} as a notebook extension.
\item We conduct a comparative user study and a one-month field study to demonstrate the effectiveness of \system{} in enhancing data flow comprehension and supporting real-world workflows.
\end{itemize}

\section{Related Work}
This section reviews studies on support for exploratory data analysis in notebooks and visualization recommendations.

\subsection{Support for Exploratory Data Analysis} 

Exploratory data analysis is inherently non-linear and iterative, yet computational notebooks typically enforce a linear cell execution structure. 
This mismatch often results in disorganized execution orders, complex cell dependencies, and multiple analysis versions, making it challenging for users to maintain awareness of the current analysis state. 
To address these challenges, researchers have developed various tools to help users manage and trace analysis stages during EDA, categorized into three types: data-centric, code-centric, and hybrid.

\textbf{Data-centric} tools focus on presenting the data patterns during EDA, including distributions and statistics. 
Some tools directly provide interfaces for visually exploring datasets~\cite{panda2025workspaceForChartCreation, voyager2}. 
Others integrate data previews into code-driven EDA environments, allowing users to examine how intermediate data states evolve.
Pandas Profiling, for example, provides detailed summaries of data features, including missing values, distributions, and correlations through histograms and heatmaps. 
Although comprehensive, these summaries can overwhelm users and require time-consuming interpretation. 
Lux~\cite{lee2021lux} and Solas~\cite{epperson2022leveraging} recommend charts to illustrate intermediate data tables. 
However, their recommended charts and their presentation order vary across cell executions, hindering consistent tracking of specific data aspects, such as attribute distributions.
AutoProfiler~\cite{autoprofiler} offers statistical summaries and distributions in a stable layout after each cell execution, updating in real-time. 
Unravel~\cite{unravel} and Xavier~\cite{xavier} further visualize how data tables change at each execution step.
However, these tools focus only on comparing the most recent data state or changes, lacking continuous tracking of data evolution throughout the complete EDA workflow.

Our approach aligns closely with these data-centric tools, sharing the goal of understanding data states during EDA. 
However, our method differs by providing continuous and consistent tracking of data throughout the entire analysis process. 
Specifically, we utilize visualizations as ``sight glasses,'' enabling users to seamlessly and comprehensively monitor global data evolution, addressing the limitations of existing tools that either lack consistency across executions or continuous data state tracking.

\textbf{Code-centric} tools focus on tracing and visualizing code dependencies and execution order, supporting management of iterative analysis workflow and multiple analytical paths. 
Unlike data-centric tools that present the effects of operations on data tables, code-centric tools primarily manage and visualize code execution structures. 
Specifically, to clarify messy code dependencies, Brown et al.~\cite{brownFacilitatingDependencyExploration2023} explore minimaps and directed graphs to visualize cell dependencies. 
Dataflow Notebook~\cite{dataflow} assigns persistent identifiers to notebook cells, explicitly encoding cell dependencies through expanded references. 
Shankar et al. introduce NBSlicer~\cite{shankar2022nbslice}, which resolves dynamic code dependencies in a bolt-on, low-overhead manner. 
Building on the dependency information, subsequent work analyzes these dependencies to improve notebook robustness and support richer notebook interaction. 
Kishu~\cite{li2024kishu} preserves notebook-state checkpoints to help users return to previous analytical contexts. 
NBSafety~\cite{macke2021nbsafe} examines fine-grained cell dependencies to detect unsafe interactions arising from out-of-order or repeated execution.
Several notebook tools~\cite{hex, pluto, marimo} further leverage dependency information to support reactive execution, automatically re-running downstream cells when their inputs change, as in Observable~\cite{observable} and IPyflow\cite{ipyflow}.
Head et al.~\cite{messnotebook} develop an extension to clean the messy notebook code. 
For managing analytical alternatives, prior work has explored alternative notebook layouts~\cite{harden2022exploring2dspace}. 
For example, Fork It~\cite{forkit} and Variolite~\cite{keryVarioliteSupportingExploratory2017} enable users to create and explore alternative code pathways. 
StickyLand~\cite{wang2022stickyland} introduces floating cells that enable users to organize code in a flexible, non-linear manner.
Verdant~\cite{verdant} records analysis activities to allow retrospective exploration of prior analysis choices. 
Recognizing the collaborative aspect of EDA, multiverse analysis approaches~\cite{multiverse, multiverseworkflow} support comparing workflows, uncertainties, and analysis variations across analysts. %
Additionally, NBSearch~\cite{nbsearch} and EDAssistant~\cite{edassistant} leverage semantic modeling to efficiently retrieve relevant scripts from code repositories. %

Although distinct from our data-focused approach, these code-centric tools provide valuable insights for managing complex code dependencies and iterative analysis paths, informing our strategy to trace data states consistently and continuously.

\textbf{Hybrid} tools integrate aspects of both code and data, primarily to manage differences across notebook versions, where both code and data outputs may vary. 
For instance, Loops~\cite{loops2025} visualizes differences across notebook versions, capturing changes in cells, scripts, and outputs, including data tables and visualizations. 
DITL~\cite{ditl} also shows the code changes through versions, but emphasizes more on the changes in data table characteristics, such as distribution shifts across notebook versions. 
However, these hybrid tools primarily highlight differences between notebook versions rather than data states within an independent analysis process, distinguishing them from the data-centric tools above and our work. 

\subsection{Visualization Recommendation}
Visualization recommendation aims to generate charts with effective visual designs and data insights based on user inputs, like a data table and optional analysis requirements. 
Existing methods can be categorized into rule-based~\cite{seedb, compassql, voyager, voyager2, apt}, learning-based~\cite{vizml, data2vis, kg4vis}, and hybrid~\cite{deepeye, deepeye2,draco} approaches. 
Rule-based methods rely on established visualization principles, such as CompassQL, which enumerates design constraints to suggest optimal charts. 
Learning-based methods~\cite{vizml, data2vis, kg4vis, vislearning, table2charts} generate visualizations end-to-end, while hybrid methods combine visualization knowledge with machine learning models. 
For instance, Draco~\cite{draco} uses answer set programming to apply visualization rules, and DeepEye~\cite{deepeye, deepeye2} employs ranking models to select the most compelling charts from a vast design space. 
Recent studies~\cite{calliope, erato, datashot, dashbot, multivision, dminer, composition, pi2} focus on generating multi-view visualizations to support complex data analysis, typically from static tables. 
In contrast, the chart recommendations in this study initiate the analysis of dynamic EDA flows.

In computational notebooks, Lux~\cite{lee2021lux} enables users to trigger visualization recommendations with a single line of code specifying intents. 
Building on Lux, Solas~\cite{epperson2022leveraging} extends recommendations by considering the history of data analysis. 
However, these methods only recommend charts for individual tables without tracking data changes across the EDA process. 
Additional support is needed to connect data insights across different cells.

\begin{figure*}[!htb]
    \centering
    \includegraphics[width=\textwidth]{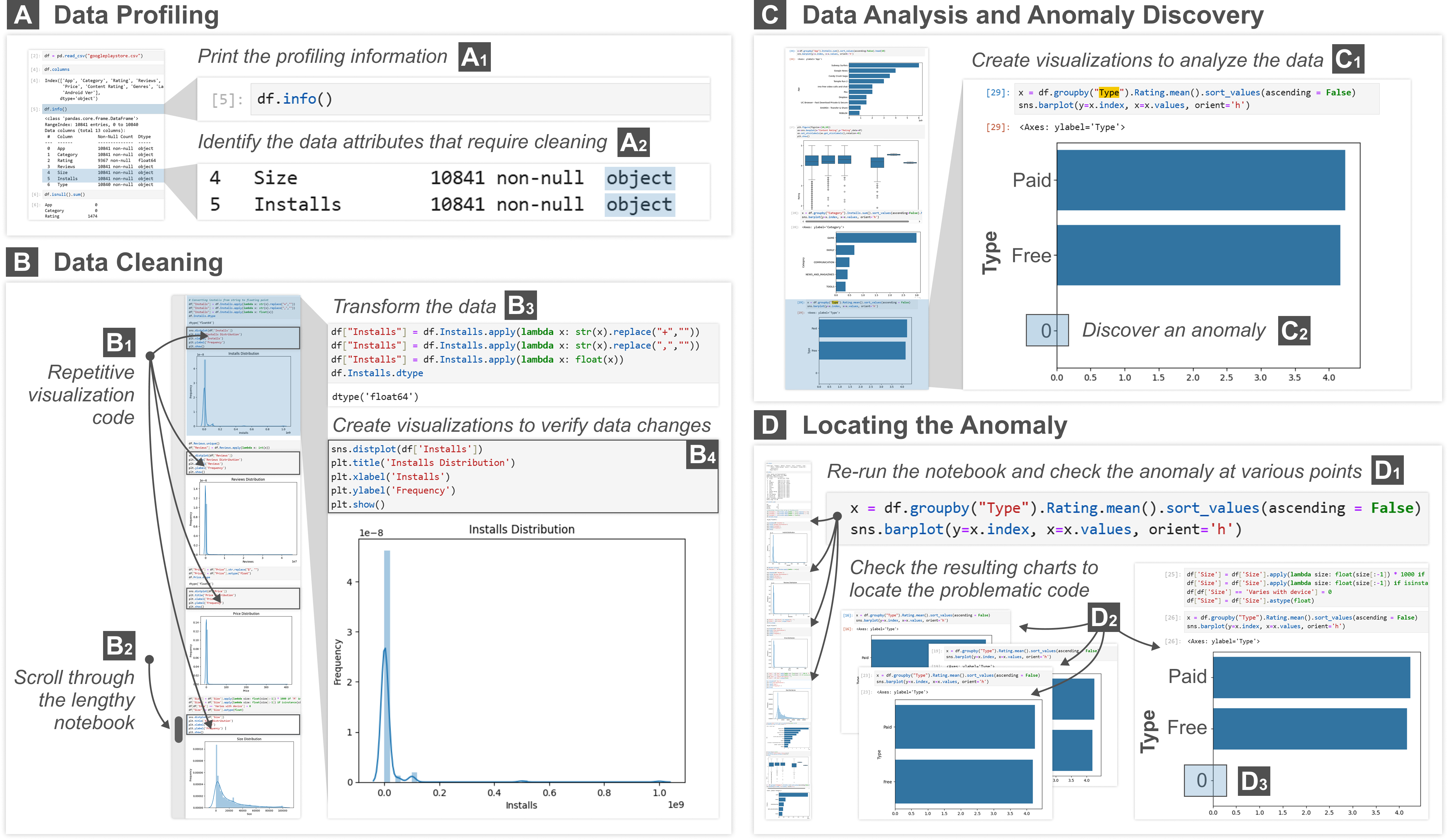}
    \caption{Motivating scenario (\autoref{sec:motivate_scenario}). 
    The EDA process will encounter a series of challenges that require tedious efforts in programming when conducting data profiling (A), data cleaning (B), data analysis and anomaly discovery (C), and locating the anomaly (D). }
    \Description{The figure presents a motivating EDA scenario that highlights tedious programming effort across four stages: data profiling, data cleaning, data analysis and anomaly discovery, and locating the anomaly. It shows that analysts often need to print profiling information, identify fields requiring cleaning, repeatedly write and rerun visualization code to verify transformations, and re-execute or scroll through a lengthy notebook to pinpoint where an anomalous result is introduced.}
    \label{fig:motivating_scenario}
\end{figure*}

\section{Problem Formulation}

We illustrate our system's motivation through a scenario that highlights how data workers monitor global data states during EDA using a basic Jupyter Notebook, along with the associated challenges they encounter. 
Following this, we review related studies to assess the progress made and the remaining gaps in current tools.
From these insights, we derive design requirements to guide the development of our system.

\subsection{Motivating Scenario}
\label{sec:motivate_scenario}

The scenario demonstrates the challenges (\textbf{CH}s) encountered while writing EDA scripts in Jupyter Notebook, where users need to monitor data states in the process. 
Alice, a data analyst, is tasked with cleaning and analyzing a dataset on Google Play Store Apps. She begins by importing the libraries and loading the dataset.

\textbf{Profiling and data cleaning. } 
After loading the dataset, Alice prints various profiling information from the dataframe (\autoref{fig:motivating_scenario}A1), such as attribute types and null value counts \textbf{(CH1)}. 
Based on the information, she identifies several attributes that require further cleaning. 
For example, string attributes like ``Installs'' and ``Size'' should be converted to numerical formats (\autoref{fig:motivating_scenario}A2). 
She then writes transformation scripts to update these attributes  (\autoref{fig:motivating_scenario}B3).
To verify the changes, she writes visualization code to inspect the distributions of the transformed attributes (\autoref{fig:motivating_scenario}B4). 
With multiple attributes requiring transformation, each involving several steps, this process quickly becomes repetitive and labor-intensive (\autoref{fig:motivating_scenario}B1), as she must repeatedly copy, edit, and run visualization scripts to monitor intermediate results \textbf{(CH1)}.
Additionally, inserting visualization scripts between transformation steps increases the overall length of the notebook (\autoref{fig:motivating_scenario}B2), making it cumbersome to navigate through the notebook and track data changes \textbf{(CH2)}. 
Alice has to scroll through the lengthy notebook to check and compare the results.

\textbf{Discover, locate, and fix the anomaly. } 
After cleaning the data, Alice begins to explore data patterns by grouping data and creating charts (\autoref{fig:motivating_scenario}C1). 
However, when she visualizes the mean ratings of free versus paid apps, she notices an unexpected zero value in the ``Type'' attribute (\autoref{fig:motivating_scenario}C2). 
Alice wonders where this issue originated and which line of scripts may have caused it. 
She first searches and confirms there are no transformation scripts directly applied to the ``Type'' attribute. 
Next, she scrolls through the notebook to identify the intermediate dataframes preceding the current one.
However, the lack of clear indications makes these intermediate states difficult to find and recognize \textbf{(CH3)}.
Moreover, as the intermediate dataframes have been overwritten, she can not create charts on them to revisit the ``Type'' attribute \textbf{(CH4)}. 
The only charts available are those created earlier, which do not include the ``Type'' attribute. 
As a result, Alice has to re-run each cell and manually insert visualization code to inspect the ``Type'' distribution (\autoref{fig:motivating_scenario}D1). 
She repeats this process until identifying the responsible code line \textbf{(CH5)}, which requires inserting and comparing multiple charts \textbf{(CH2)}.
By comparing the charts step by step (\autoref{fig:motivating_scenario}D2), she discovers that the issue is caused by an operation applied to the ``Size'' attribute, which mistakenly sets many data rows to zero (\autoref{fig:motivating_scenario}D3).
After fixing the code, Alice reruns the subsequent cells to verify the correction. 
However, as the cells are re-executed, the previous charts are overwritten, forcing her to compare the results mentally, which makes it difficult to track the changes before and after the fix \textbf{(CH4)}.

In summary, monitoring data states during EDA in traditional notebook workflows faces the following challenges:
\begin{enumerate}[leftmargin=*, label={\textbf{CH\arabic*.}}]

\item \textbf{Tedious manual programming for data visualization.}
Users have to manually create and update visualization scripts, which is both time-consuming and error-prone.

\item \textbf{Cognitive overload from comparison among cells.} 
Understanding how data evolves requires comparing results across multiple cells. 
As the notebook grows and the number of cells increases, mentally remembering and comparing these results becomes increasingly demanding.

\item \textbf{Lack of clear indication in the relationships between dataframes.}
Users struggle to identify where a dataframe originated and how it was transformed throughout the analysis, as these relationships are buried in the script and not visually represented. 

\item \textbf{Loss of execution history and previous data states.}
During EDA, previous states of data may be overwritten, and the history of re-executed cells may be obscured, making it difficult to revisit and compare data states over time.

\item \textbf{Repeatedly copying and editing the same visualization codes.}
To monitor the data states in specific attributes, users repeatedly copy and paste the same visualization code, which is inefficient and redundant.

\end{enumerate}

\subsection{Progress and Gaps in Current Tools}
Several studies have attempted to enhance EDA by providing visualizations that help users better understand data states (\textbf{CH1}). 
The most notable examples are Lux~\cite{lee2021lux}, Solas~\cite{epperson2022leveraging}, and AutoProfiler~\cite{autoprofiler}.
Lux and Solas recommend visualizations based on data insights and analysis history, offering a quick overview of patterns and trends that align with the user's operations and specified intents. 
AutoProfiler focuses on displaying basic information about the data, such as the distribution and profiling of each attribute. 
We consider both data insights and profiling valuable and incorporate these factors into our chart recommendations.

On the other hand, both Lux and AutoProfiler primarily focus on local data states and lack support for understanding the global state of the data (\textbf{CH2-CH5}). 
Lux requires users to write specific visualization intent codes to display the current state of a dataframe. 
To monitor global state changes, users must repeatedly insert the codes, which only simplifies the process of writing visualization codes compared to traditional workflows. 
AutoProfiler eliminates the need for code insertion by automatically updating visualizations and profiling to reflect the most recent state. 
However, it only shows the latest state of the data and does not support revisiting or comparing previous states.
In this study, we enable users to use recommended charts as ``sight glasses'' to trace the global data flow, enabling users to track and understand how data evolves throughout the entire EDA process.

\subsection{Design Requirements}

\begin{figure*}[!htb]
    \centering
    \includegraphics[width=\textwidth]{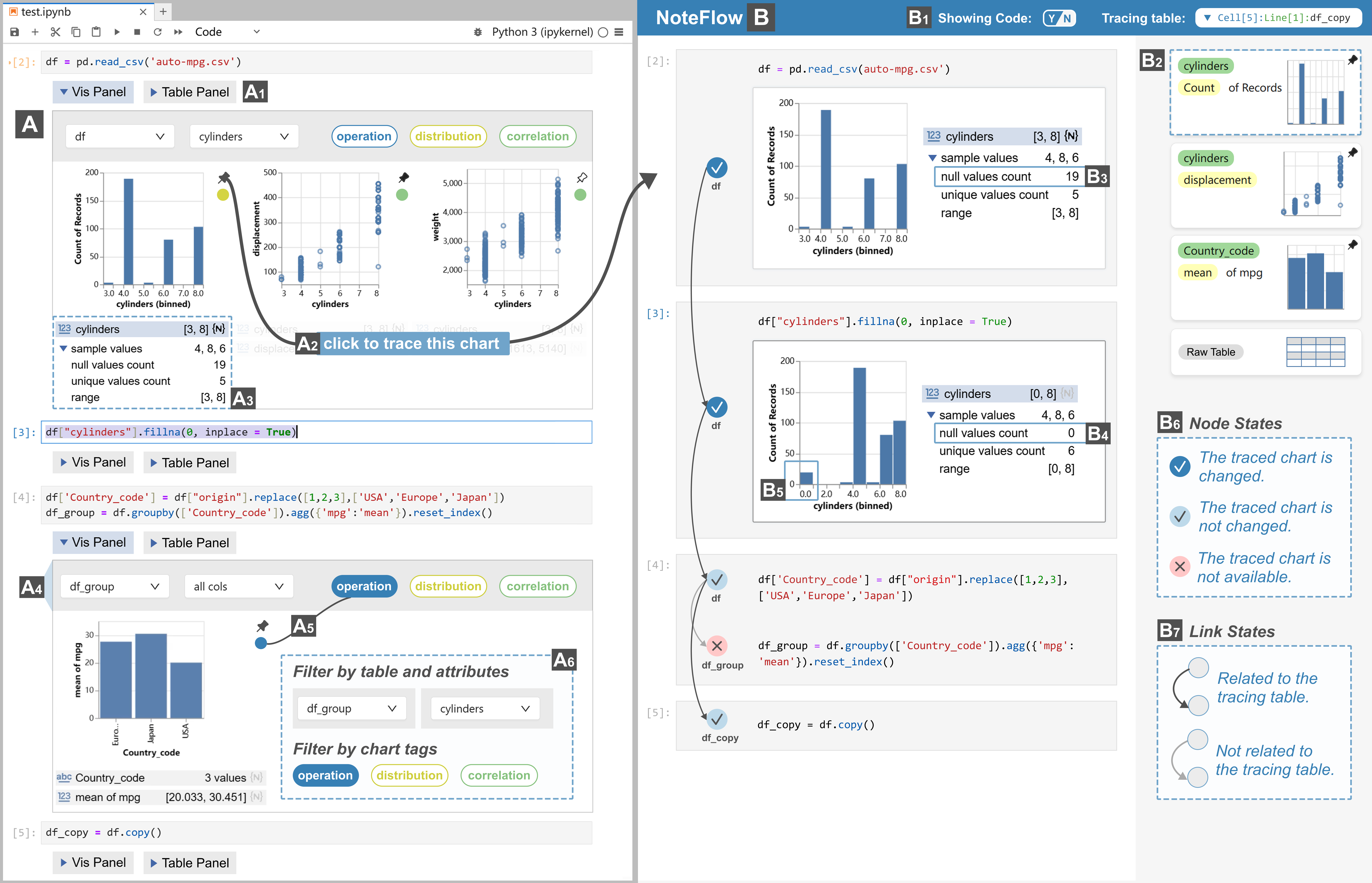}
    \caption{
        The interface of \system{} (\autoref{sec:system}), consisting of a chart view (A) and a flow view (B).
        The chart view displays recommended charts under each cell, each paired with profiling information of the corresponding attributes (A3).
        Users can filter charts by dataframe, attribute, or tag (A4, A6), and freely select any chart of interest (A2) to trace its evolution.
        The selected chart is then visualized in the flow view (B), which shows how the chart changes across the data flow and highlights when it is updated or unavailable (B6).
        \system{} also maintains a list of all tracing charts, allowing users to switch between them (B2).
        }
        \Description{The figure shows the NoteFlow interface with two coordinated views: a chart view and a flow view. The chart view presents recommended charts under each notebook cell, each accompanied by profiling information for the visualized attributes, and supports filtering by dataframe, attribute, or chart tag. Users can select any chart to trace, after which the flow view displays that chart across the data flow, indicating where it changes or becomes unavailable. The system also keeps a list of traced charts for switching among them.}
    \label{fig:interface}
\end{figure*}

Based on the challenges identified in the motivating scenario and the progress and gaps in current tools, we derive the following design requirements (\textbf{RQ}s) to guide our system's development:

\begin{enumerate}[leftmargin=*, label={\textbf{RQ\arabic*.}}]

    \item \textbf{Effective chart recommendation.}
    The system should recommend charts that effectively represent the key aspects of the data, considering both data insights and profiling information. These recommendations should help users quickly grasp data states at any given point, reducing the need for manual chart creation.

    \item \textbf{On-demand tracing of changed related data.}
    The system should allow users to trace the data flow on demand, enabling them to focus specifically on the related intermediate dataframes where changes have occurred. This targeted approach helps users concentrate on critical transformations without being overwhelmed by irrelevant information.
    
    \item \textbf{Representation of data relationships.}
    Given the complexity and number of intermediate dataframes in a notebook, it is essential to provide visual cues that clearly indicate the flow between these dataframes. 
    This flow should include information about the transformations applied, such as \code{df\_2} being generated from \code{df\_1} through filtering. These visual hints will help users quickly understand how data evolves throughout the notebook.
    
    \item \textbf{Support for revisiting past intermediate dataframes.}
    The system should provide mechanisms to revisit past intermediate dataframes, even if they have been overwritten or hidden due to cell re-execution. This feature is crucial for allowing users to compare historical data states with current ones, ensuring that they can accurately track changes and understand the impact of each transformation. 

    \item \textbf{Consistent tracing of charts.} 
    To facilitate a comprehensive understanding of how data changes over time, the system should support the consistent tracing of data states. This involves tracking the changes in specific charts or visual representations as they reflect the underlying data across multiple transformations. 
\end{enumerate}

\section{\system{}}
\label{sec:system}
In this section, we introduce the interface of \system{} and explain how it addresses the previously identified requirements.
\system{} is implemented as an interactive extension integrated into Jupyter Notebook to support consistent and continuous tracing of data flow. 
Specifically, our current implementation targets the Pandas library, handling data tables stored in Pandas dataframes and transformations expressed as Pandas operations. 
The interface consists of two main components: the chart view (\autoref{fig:interface}A) and the flow view (\autoref{fig:interface}B).

\subsection{Chart View}
The chart view (\autoref{fig:interface}A) is placed under each cell, presenting recommended charts for the dataframes generated within that cell.
It allows users to explore and filter these charts to better understand the local data states (\textbf{RQ1}).
To avoid information overload, the chart view is collapsed by default and can be expanded by clicking the ``Vis Panel'' button under each cell (\autoref{fig:interface}A1).
\autoref{fig:interface} shows two examples of expanded chart views, with the button color changed to indicate their active state.
Each expanded chart view contains two parts: a filtering panel (\autoref{fig:interface}A4) and a ranked list of recommended charts below it.

\textbf{Filtering panel.}
The filtering panel allows users (1) to select the dataframe and one or more attributes to scope the recommendations, and (2) apply chart tags of interest to narrow the results (\autoref{fig:interface}A6).
Since a cell may create or update multiple dataframes, users can specify which one to view its associated charts.
By default, \system{} displays charts for the most recently updated dataframe. 
For example, in \code{cell 4}, the latest dataframe is created on \code{line 2} and named \code{df\_group}, so the recommended charts are derived from this dataframe (\autoref{fig:interface}A4).
Users can further refine results by selecting attributes via dropdowns.

Users can also filter results using chart tags, including operation (charts related to recent operations and operated attributes), distribution (attribute histograms), and correlation (scatter plots of attribute pairs).
In the example shown, the ``operation'' tag is selected (\autoref{fig:interface}A5), resulting in a chart that visualizes the average \code{mpg} grouped by \code{Country\_code}, reflecting the effect of the \code{groupby} operation in \code{line 2} of \code{cell 4}.
The chart list updates dynamically as filters are applied, helping users efficiently locate charts that match their needs (\textbf{RQ1}).

\textbf{Chart recommendations.}
Recommended charts are ordered with those involving the operated attributes ranked first (see~\autoref{sec:ranking} for ranking details), helping users quickly inspect the impact of their latest operations.
Each chart is accompanied by attribute profiling details (\autoref{fig:interface}A3) to support interpretation, and relevant tags (e.g., operation, distribution, correlation) are shown as colored circles (\autoref{fig:interface}A5) to indicate whether the chart reflects recent operations or highlights notable data facts.
All historical dataframes in executed cells are preserved during the analysis, allowing users to revisit them at any time (\textbf{RQ4}). 

Charts can also be pinned (\autoref{fig:interface}A1) to trace the corresponding data flow with consistent visual encodings in the flow view (\autoref{fig:interface}B). This tracing can be applied to any chart from any dataframe in any executed cell, allowing on-demand tracking of attributes of interest (\textbf{RQ2}).

\begin{figure}[!htb]
    \centering
    \includegraphics[width=\linewidth]{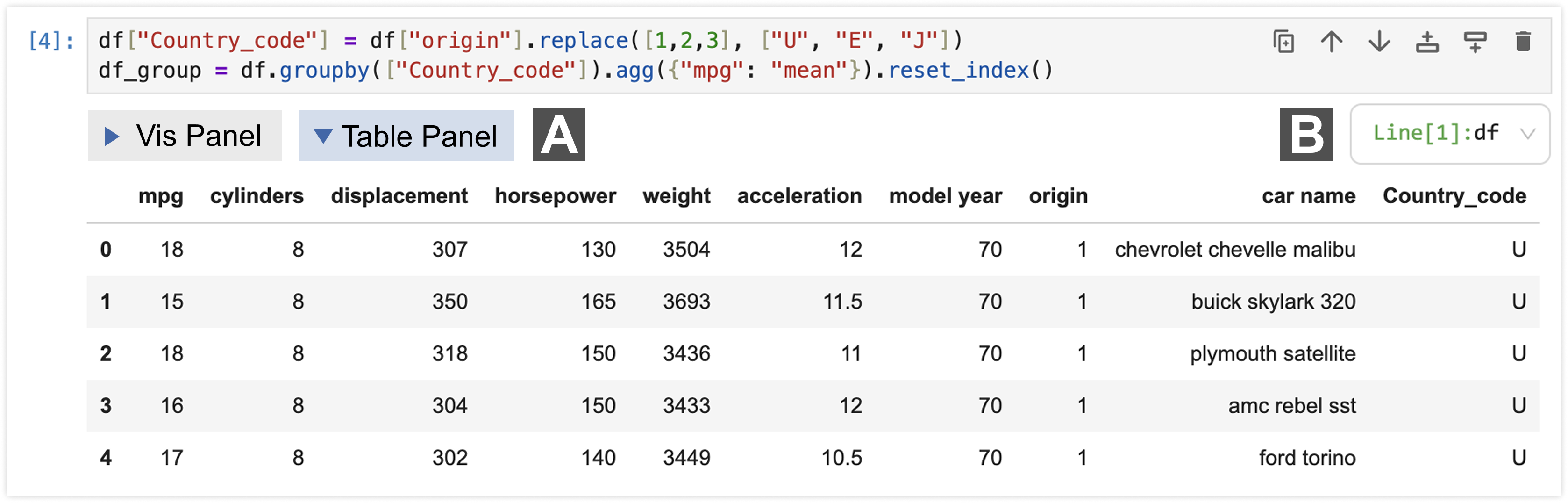}
    \caption{Table panel of the chart view. Users can expand the panel (A) to preview the first five rows of the specified dataframe (B).}
    \Description{The figure shows the table panel in NoteFlow's chart view. Users can expand the panel to preview the first five rows of a selected dataframe, and choose which dataframe to display using a dropdown menu.}
    \label{fig:table_panel}
\end{figure}

In addition to the filtering panel and chart recommendations, the chart view includes a table panel (\autoref{fig:table_panel}) that displays the first five rows of a selected dataframe, enabling quick inspection of raw records. Similar to the visualization panel, users can click to expand the table panel and switch between dataframes as needed.

\begin{figure*}[!htb]
    \centering
    \includegraphics[width=.8\textwidth]{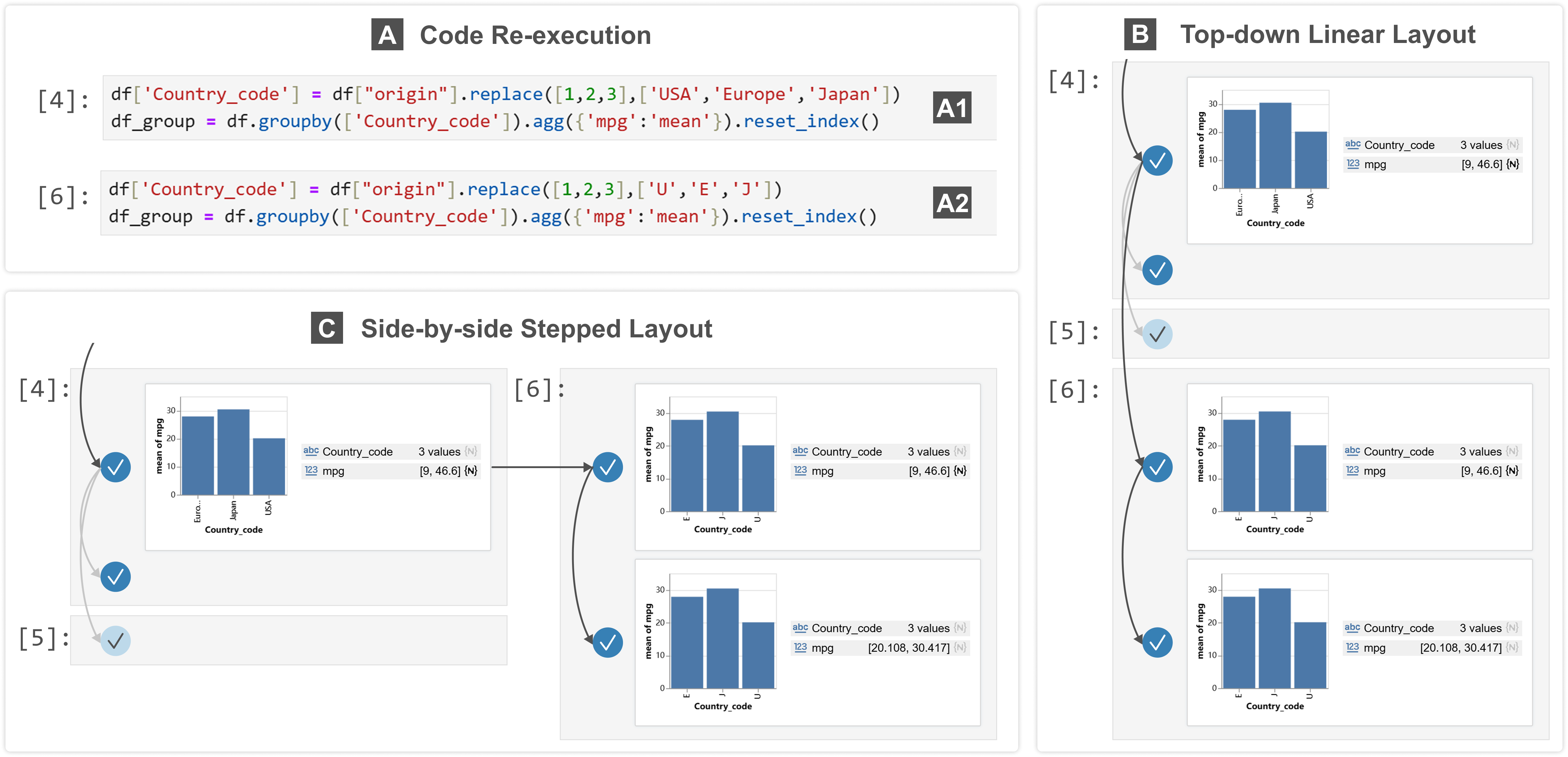}
    \caption{Design alternatives for re-running cells.
    (A) Code re-execution: the analyst modifies and re-runs \code{cell 4}, producing a new execution \code{cell 6}. 
    (B) Top-down linear layout (default): nodes are arranged strictly in execution order from top to bottom. 
    (C) Side-by-side stepped layout (alternative): different versions of the same cell are displayed next to each other.}
    \Description{The figure compares design alternatives for handling re-running notebook cells in NoteFlow. It shows an analyst modifying and re-executing a cell, creating a new version that produces updated results. Two layout options are illustrated: a default top-down linear layout that orders nodes strictly by execution order, and an alternative side-by-side stepped layout that places different versions of the same cell next to each other.}
    \label{fig:re-run}
\end{figure*}

\subsection{Flow View}
The flow view (\autoref{fig:interface}B) provides an overview of the whole EDA flow with the consistent traced chart (\textbf{RQ5}).
It consists of a node-link diagram with traced charts and a tracing chart list.

\textbf{Node-link diagram with traced charts. }
The node-link diagram visualizes the lineage of dataframes across notebook cells (\textbf{RQ3}). 
Each grey block corresponds to a cell, labeled with its execution number consistent with the notebook interface. 
Within each cell, \textbf{nodes} represent intermediate dataframes produced after each line of code in that cell. 
Each node is accompanied by the corresponding code snippet, together with the traced chart and attribute profiling details.
\textbf{Links} between nodes represent input–output relationships between dataframes across code lines. 
When a chart is selected for tracing (\autoref{fig:interface}A2), its encoding is propagated to all nodes. 
This shows before-and-after dataframe states using a consistent visual encoding, enabling users to compare dataframes across the notebook and observe how the data evolves. 
This diagram is dynamically updated as new cells are executed.

\textit{Nodes. }
Specifically, nodes are color-coded to indicate tracing states (\autoref{fig:interface}B6).
\textbf{Dark blue} nodes mark code lines where the traced chart differs from the previous state and are expanded by default to support direct comparison (\textbf{RQ4, RQ5}).
The initial chart, which has no prior state, is also shown in dark blue.
For example, nodes in \code{cell 2} and \code{cell 3} are both dark blue and expanded. 
The histogram and profile of \code{cylinders} in \code{cell 3} show differences from \code{cell 2}, where filling missing values introduced additional zeros (\autoref{fig:interface}B3–B5).
\textbf{Light blue} nodes indicate no change in the traced chart and are collapsed by default. 
For example, the first node in \code{cell 4} and the node in \code{cell 5} do not alter the distribution of \code{cylinders}, so their charts remain unchanged and are hidden. 
Users can also manually expand or collapse nodes by clicking on them.
\textbf{Red} nodes represent untraceable states, where the attribute required for rendering the traced chart no longer exists. For instance, in \code{cell 4}, the \code{groupby} operation removes \code{cylinders}, preventing the chart from being rendered.

\textit{Links. }
Since code may branch into multiple outputs, link colors clarify whether the branch is related to the traced chart: black links indicate transformation lineage of the traced dataframe, while grey links denote branches not directly involved in the trace (\autoref{fig:interface}B7). 
For example, when tracing the dataframe \code{df\_copy}, the path from \code{cell 2}'s \code{df} to the \code{df\_copy} in \code{cell 5} is marked with black links, whereas the unrelated \code{df\_group} in \code{cell 4} is connected by grey links. 
By default, tracing follows the lineage of the latest dataframe in the most recently executed cell, and users can change the base table via the dropdown menu (\autoref{fig:interface}B1).

\textbf{Tracing chart list.}
All traced charts are displayed in the right panel (\autoref{fig:interface}B2), where users can switch charts to trace by selecting a chart card. 
In the example shown, the first chart (count of records by cylinders) is selected and displayed throughout the flow. 
Users may also choose to trace the raw dataframe, in which case each node is displayed as a table preview, allowing them to identify where the dataframe changed and how it propagated.

\textbf{Top-down linear layout for re-running cells. }
\system{} supports tracing tables even when cells are re-executed, utilizing a top-down linear layout. 
As shown in~\autoref{fig:re-run}A, the analyst revises and re-runs \code{cell 4}, producing a new result in \code{cell 6}. 
In the flow view (\autoref{fig:re-run}B), all nodes are arranged strictly in execution order from top to bottom, providing a clear global context of execution history. 
This linear layout helps analysts follow the true order of operations in notebooks, which otherwise lack an explicit execution sequence due to out-of-order runs. 
The solid black links indicate dependencies, allowing users to see how each table connects to its predecessors and successors.

In an earlier version, we implemented a side-by-side stepped layout for re-running cells (\autoref{fig:re-run}C), where a new node \code{[6]} was placed to the right of node \code{[4]} so that old and new results could be viewed together.
While this layout enabled side-by-side comparison between different execution versions, it consumed substantial space and scaled poorly with repeated re-runs.
During the field study (\autoref{subsection:field_study_rerun}), both layouts were provided, and data scientists expressed a preference for the linear layout in real workflows. 
Consequently, we adopted the linear layout as the default design.

\section{Method}
\label{sec:method}

This section describes the construction of \system{}.

\subsection{Framework Overview}

\begin{figure}[!htb]
  \centering
  \includegraphics[width=\linewidth]{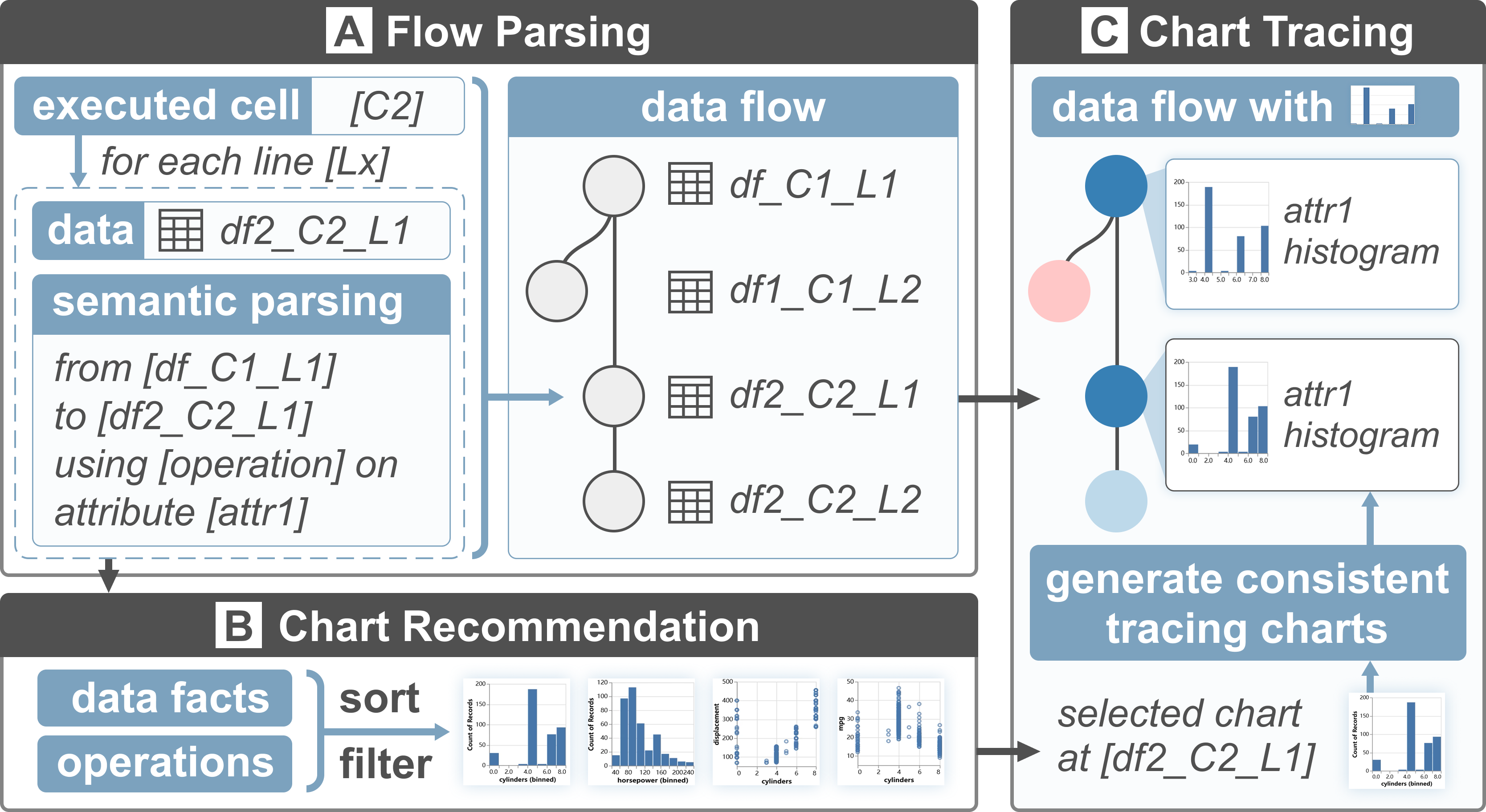}
  \caption{
  	The Framework of \system{}, including three modules: flow parsing (A), chart recommendation (B), and chart tracing (C). 
    }
    \Description{
    The figure summarizes NoteFlow's framework with three modules. In flow parsing, the system parses each executed cell line by line, identifies intermediate dataframes, and links input and output tables through the transformation to construct a data-flow graph. In chart recommendation, it ranks and filters candidate charts based on data facts and operations. In chart tracing, given a user-selected chart, the system generates consistent charts along the data flow so the same visual perspective can be compared across related tables.
    }
  \label{fig:framework}
\end{figure}

To effectively monitor the evolution of data throughout the EDA process, \system{} is built around three modules: flow parsing, chart recommendation, and chart tracing (\autoref{fig:framework}). 
Together, these three components form our conceptual framework for supporting data flow tracing during the analysis.

\textbf{Flow Parsing.}
Data flow parsing aims to extract the dataframes and the relationships between them.
For each cell execution, NoteFlow parses every code statement to identify the input dataframes, the resulting output dataframes, the applied operations, and the attributes involved.
Together, these elements form a dynamic data flow graph in which each dataframe state is a node, and edges correspond to operations that connect input dataframes to the resulting output dataframes.
Intermediate dataframe states are preserved, providing an evolving record of the analysis. 
This parsed flow serves as the foundation for chart recommendations and chart tracing.

\textbf{Chart Recommendation.}
This module generates recommended charts under each executed cell, with each chart also serving as an entry point for chart tracing. 
Building on the intermediate dataframes preserved by flow parsing, it generates visualizations that capture data transformations and data insights. 
From the perspective of data changes, charts highlight how operations affect the dataframe (e.g., distributions before and after a transformation). 
From the perspective of data insights, charts surface patterns such as value distributions or correlations. Together, these recommendations allow users to quickly assess dataframe states without manually writing visualization code.

\textbf{Chart Tracing.}
Building on flow parsing and chart recommendation, chart tracing enables users to explore visually consistent charts across the entire data flow. 
Starting from a recommended chart for a specific dataframe produced by a cell execution, the system propagates its encoding to all nodes along the corresponding lineage, revealing where the chart has changed, remained unchanged, or become unavailable. 
To achieve this, \system{} (1) adapts traced chart encodings to both remain consistent and reflect dataframe transformations, ensuring that visual representations remain coherent and meaningful, 
and (2) determines, from the perspective of the chart's related attributes, whether the state represents a change, no change, or an unavailable case.

The following subsections explain the details of each module.

\subsection{Flow Parsing}
We first parse and extract the data flow, which serves as the foundation for subsequent chart recommendation and chart tracing. 
The general data flow in computational notebooks can be highly complex, involving diverse data formats, multiple library functions, and user-defined functions.
In this study, we scope the problem to data flow consisting of Pandas dataframes and Pandas operations, as Pandas is one of the most widely used libraries for EDA.

Considering the data flow as a graph, each node represents a snapshot of a dataframe, and edges correspond to operations that connect input dataframes to the resulting output dataframe. 
To construct this graph, NoteFlow parses each code statement using AST analysis to identify the input and output dataframes, invoked Pandas functions, parameters, and the attributes involved.
NoteFlow then determines transformation operation types based on the functions and parameters following the taxonomy of Comantics~\cite{comantics}. 
Specifically, for each code statement, we:
\begin{itemize}[leftmargin=*]
    \item Identify whether the output belongs to Pandas dataframes.
    \item If so, extract the variable names of input and output dataframes, the Pandas function names, and their parameters.
    \item Forward function names and parameters to a detection module that identifies transformation operations based on prior work~\cite{comantics, proactive, tablescraps}, and parse the parameters to identify the attributes involved in the operation.
\end{itemize}
Each statement is represented as a list of tuples, with each tuple corresponding to one output dataframe and defined as (input dataframes, output dataframe, operations, involved attributes). 
A single statement may produce multiple output dataframes, which in turn generate multiple tuples, corresponding to multiple nodes. 
Within each tuple, there may be multiple operations (e.g., chained calls such as \code{fillna().groupby()}) and multiple input dataframes, resulting in multiple source nodes and links. 
All of these cases are incorporated into the construction of the node-link graph.

We also account for several special cases during parsing. 
First, if a function produces a dataframe without being assigned to a variable, we still capture its output as a node in the data flow, since notebooks display such results by default. 
Second, iterative code, such as loops, may generate many intermediate dataframes. 
To avoid clutter, we preserve only the first and last iterations by default. 
Finally, for user-defined functions, we do not trace inside the function bodies and thus cannot identify transformation types, but we record their inputs, outputs, and attributes passed as parameters, which still allows us to capture the dataframe state before and after execution.

The final result is a graph of nodes and edges representing the parsed data flow. Each node corresponds to a snapshot of a dataframe's state, not the variable itself, since a variable may change over time. For example, in the code \code{df = df + 1}, the variable \code{df} is represented by two distinct nodes: one before the addition and one after. Internally, each node is uniquely identified by the variable name, the cell execution ID, and the line ID of the generating statement. 
A detailed illustration of this construction is provided in the supplementary materials.

\begin{table*}[!htb]
  \caption{Chart Recommendation Rules}
  \small
  \label{tab:rules}
  \centering
  \begin{tabular}{p{2cm}p{4.5cm}p{2cm}p{5cm}}
    \toprule
    Factors&Description&Chart Type& Factor Values\\
    \midrule
    Transformation Operations & The chart is suitable for showing specific attribute value changes after the operation. & histogram, bar, line, heatmap& mutate, filter, aggregate, sort, fill, replace, unfold, extract, deduplicate, fold, separate, merge\\
    \midrule
    Data Facts & The chart is suitable for showing the data facts of specific attributes. & histogram, line, scatter&distribution, trend, correlation\\
  \bottomrule
\end{tabular}
\end{table*}

\subsection{Chart Recommendation}
\label{sec:ranking}

After each cell execution, \system{} recommends charts that highlight both \emph{data changes} introduced by operations and \emph{data insights} inherent in the dataframe. The recommendation process consists of two phases: querying and ranking. In the querying phase, charts are generated based on a set of rules, while in the ranking phase, the generated charts are ordered according to predefined criteria.

In the querying phase, charts are generated based on transformation operations (to reflect data changes) and data facts (to reveal insights).
Prior work such as Lux~\cite{lee2021lux} recommends charts by detecting data facts, while Solas~\cite{epperson2022leveraging} further considers transformation operations.
Inspired by these approaches, \system{} employs a similar approach that integrates both perspectives by considering operations and data facts together. 

\textbf{Transformation operations.}
EDA often involves reshaping dataframes through transformation operations, and recommended charts should capture how these operations alter or create attributes.
Previous studies~\cite{tablescraps, comantics, proactive} have presented different data operation taxonomies. 
Specifically, Xiong et al.~\cite{comantics} introduced a comprehensive set of 30 categorized transformation operations.
Building upon this foundation, we focus on whether operations modify attribute distributions or introduce new attributes, and identify 12 target operations (\autoref{tab:rules}).

Following visualization design principles~\cite{draco}, which link visualization tasks and data types to appropriate visual encodings, we treat each transformation operation as a visualization task whose goal is to reveal its effect on the resulting dataframe. 
For each operation, we design rules that select chart types and encodings based on the task implied by that operation and the data types involved.
For example, filling \code{NA} changes the distribution of an attribute, which is best examined through a histogram.
For \code{groupby-aggregate} operations, the grouping attribute naturally serves as the x-axis and the aggregated values (e.g., mean of a numeric attribute) as the y-axis, enabling direct inspection of group-level patterns.
These rules form the basis for generating charts that reveal data changes in the resulting dataframes.

\textbf{Data facts.} 
We consider three types of data facts: distribution, correlation, and trend, and visualize these data facts with appropriate chart types (\autoref{tab:rules}). 
For distribution visualization, we derive histograms that show the data distribution of a specific attribute.
For correlation visualization, we visualize the attribute pairs using scatterplots.
For trend visualization, we show the trend of an attribute by a temporal attribute using a line chart.
Since correlations can generate a large number of attribute pairs, we limit recommendations to the top 10 pairs with the highest correlation coefficients when more than 20 attributes are present. 
If users want additional scatterplots, they can create a dataframe containing only the relevant attributes (fewer than 20), in which case correlation plots for all of those attributes will be shown.

\textbf{Chart ranking.}
The querying phase can generate many charts, and a single chart may be triggered by multiple reasons (e.g., a histogram that reflects both an operation and a distribution fact). 
Such charts are tagged with all applicable reasons. 
We then rank charts according to the following criteria (highest to lowest priority):
\begin{itemize}[leftmargin=*]
\item Relevance to recent operations: charts that directly reflect the operations are prioritized.
\item Distribution facts: charts showing attribute distributions are ranked higher, as they support intuitive profiling.
\item Inclusion of operated attributes: charts involving attributes directly modified by operations (e.g., renamed, filtered, or selected) are prioritized.
\item Number of tags: charts tagged with more recommendation reasons are ranked higher.
\item Correlation strength: scatterplots with stronger correlation scores are prioritized.
\end{itemize}

\subsection{Chart Tracing}

\system{} enables users to monitor data state evolution by tracing recommended charts across the data flow.
Users can select any recommended chart from any executed cell, after which \system{} applies the consistent chart encoding to all intermediate dataframes.
This allows users to observe how the chart evolves and identify where changes occur.
The mechanism relies on two key components: chart encoding adaptation and attribute change detection.

\textbf{Chart encoding adaptation. }
Due to the flexibility of data operations, chart encodings may not always be applicable to all dataframes, particularly when attributes are added or deleted. 
For example, tracing a histogram of attribute \code{B} backward is impossible if \code{B} was created by mutating \code{A}, since \code{B} does not exist before the mutation. Similarly, deleting an attribute breaks forward tracing, while combinations of additions and deletions can disrupt tracing in both directions.

To address these cases, \system{} strives to maintain tracing relationships as consistently as possible. When a traced chart's encoding becomes inapplicable, the system automatically detects the operation's input–output relationships and substitutes the traced attribute with its corresponding input or output. 
For instance, if a histogram of attribute \code{B} is traced and \code{B} is derived from \code{A}, the system substitutes it with a histogram of \code{A} before the mutation.
Following the categorization of Xiong et al.~\cite{comantics}, \system{} identifies transformation operations that may add or delete attributes. Of the 30 transformation types, 15 have the potential to affect tracing. By recognizing these cases and adapting the tracing process accordingly, \system{} ensures consistent chart tracing throughout data evolution.

\textbf{Attribute change detection.}
When the traced chart's attributes remain present in a dataframe, \system{} further determines whether their underlying values have changed to decide the node's tracing state (i.e., whether to expand or collapse it in the flow view). 
The detection combines lightweight statistics and robust hashing. 
Specifically, we first compare basic properties of the attribute, including length, number of missing values, and data type. 
For numerical and temporal attributes, we also check minimum and maximum values. 
If these properties remain unchanged, we then compute a content-based hash of the attribute and aggregate it with a bitwise XOR to test equivalence. 
This information drives the expansion and collapsing of nodes in the flow view, allowing users to focus on and locate where meaningful data changes occur.

\section{User Study}
To evaluate the effectiveness and usability of \system{}, we conducted a user study comparing our system with Lux~\cite{lee2021lux}. 
This study was designed to understand whether and how \system{} can help users better monitor the evolving data flow during EDA. 

\subsection{Baseline Selection: Lux}
We compare \system{} with Lux~\cite{lee2021lux}, a state-of-the-art chart recommendation tool for Jupyter Notebooks.
Lux augments Pandas outputs with recommended visualizations and supports custom chart intents, helping users quickly inspect the current data state.

We chose Lux based on the following considerations.
As \system{} focuses on tracing data flow during EDA, we sought a baseline that supports effective inspection of evolving data states.
While existing tools offer code or version tracking~\cite{forkit,brownFacilitatingDependencyExploration2023,loops2025}, they do not visualize data states.
Among data-centric tools~\cite{lee2021lux, epperson2022leveraging,autoprofiler}, Lux stands out as the most popular chart recommendation tool for computational notebooks, and offers basic support for revisiting past data states through recommendation history.
As such, we consider Lux appropriate as our baseline.

\subsection{Participants} 
We recruited 12 data analysts from diverse fields, including sports science, computer science, and data science, comprising four female and eight male participants. 
On average, participants have 4.3 years of experience in data analysis and 3.6 years of experience using computational notebooks. 
All participants have extensive experience in monitoring data states by creating data profiles and charts during their analyses.

\subsection{Task Design}

\textbf{Overview.}
Participants were provided with two pre-written data analysis notebooks, each paired with a dataset. 
Their task was to understand the data flow in the notebooks using \system{} and Lux, respectively.

To better quantify the effects of data flow understanding, each notebook was embedded with an anomaly in the transformed data table caused by an intermediate operation. 
Participants were informed that each notebook included such an anomaly and were required to identify and locate the precise line of code responsible for the anomaly. 
They needed to provide specific evidence (e.g., tables, charts, or other presented information) to justify their findings, demonstrating an understanding of the data flow beyond solely reading the code. 
During the task, participants were free to inspect, modify, and execute the code cells as needed. 
The two notebooks, associated datasets, and embedded anomalies are as follows. 

\begin{itemize}[leftmargin=*]
    \item \textbf{N1: Google Play Store Apps.} 
    This notebook involved a dataset about Google Play Store apps. 
    An anomaly was introduced by a transformation that incorrectly sets all rows to zero where the ``Size'' attribute equaled ``Varies with device,'' which later leads to an unexpected zero value in the ``Type'' attribute.
    \item \textbf{N2: COVID-19 Data.} 
    This notebook focused on a dataset related to COVID-19, where several countries' recovery data was null after 2021. 
    The anomaly was introduced through a transformation that dropped all rows with null values, resulting in missing data for certain countries post-2021. 
\end{itemize}

To reduce cognitive load, we kept the notebooks concise while retaining a complete EDA pipeline, including data loading, profiling, cleaning, and analysis. 
N1 and N2 contained 8 and 7 cells, respectively, averaging 1.75 and 2.71 lines of code per cell.
The datasets and notebooks are included in the supplementary materials.

\textbf{Anomaly details.}
The anomalies in both notebooks were caused by a specific line of intermediate transformation, both belonging to common types of real-world data analysis errors~\cite{yangSubtleBugsEverywhere2021}. 
These transformations often appear correct when read in isolation (e.g., processing special values in N1 or dropping nulls in N2). 
However, they can introduce issues that are not immediately apparent without comprehensively inspecting data states.
This design required participants to reason through the data flow and understand the relationships between different dataframes, rather than relying on surface-level code reading. 
It thus provided a rigorous test of each system's ability to support data flow tracing.

Moreover, we exposed the anomalies in the analysis stage, near the end of each notebook. 
This served two purposes: 
(1) it mirrored real-world scenarios where data analysts often discover problems during final analysis and must trace back to earlier steps, 
and (2) it kept task duration manageable, while still demanding that participants engage with tracing the full data flow.

\subsection{Procedure} 
All participants began the study by completing a consent form and a demographic questionnaire covering age, education, occupation, and relevant background.
Next, participants were given a brief introduction to the study's background and tasks (5 minutes).

\textbf{Formal study (about 40 minutes): }
The formal study comprised two sessions, each focusing on one of the systems.
Participants were assigned to one of the two systems (Lux or \system{}) and one of the notebooks (N1 or N2) for the first session, and then switched to the other system and notebook for the second session.
To ensure fairness and mitigate order effects, the assignment and orders of notebooks and systems were fully counterbalanced. 
Specifically, each session began with an 8-minute introduction to the system, where participants were guided through the system's features and interactions using a notebook and a dataset different from those used in the tasks. 
After the introduction, participants were given 5 minutes to explore the system freely.
Participants then proceeded to complete the task introduced above using the specified system and notebook with a maximum time limit of 8 minutes.
During the process, participants followed a think-aloud protocol, and all their actions were recorded.

\begin{figure}[!htb]
    \centering
    \includegraphics[width=\linewidth]{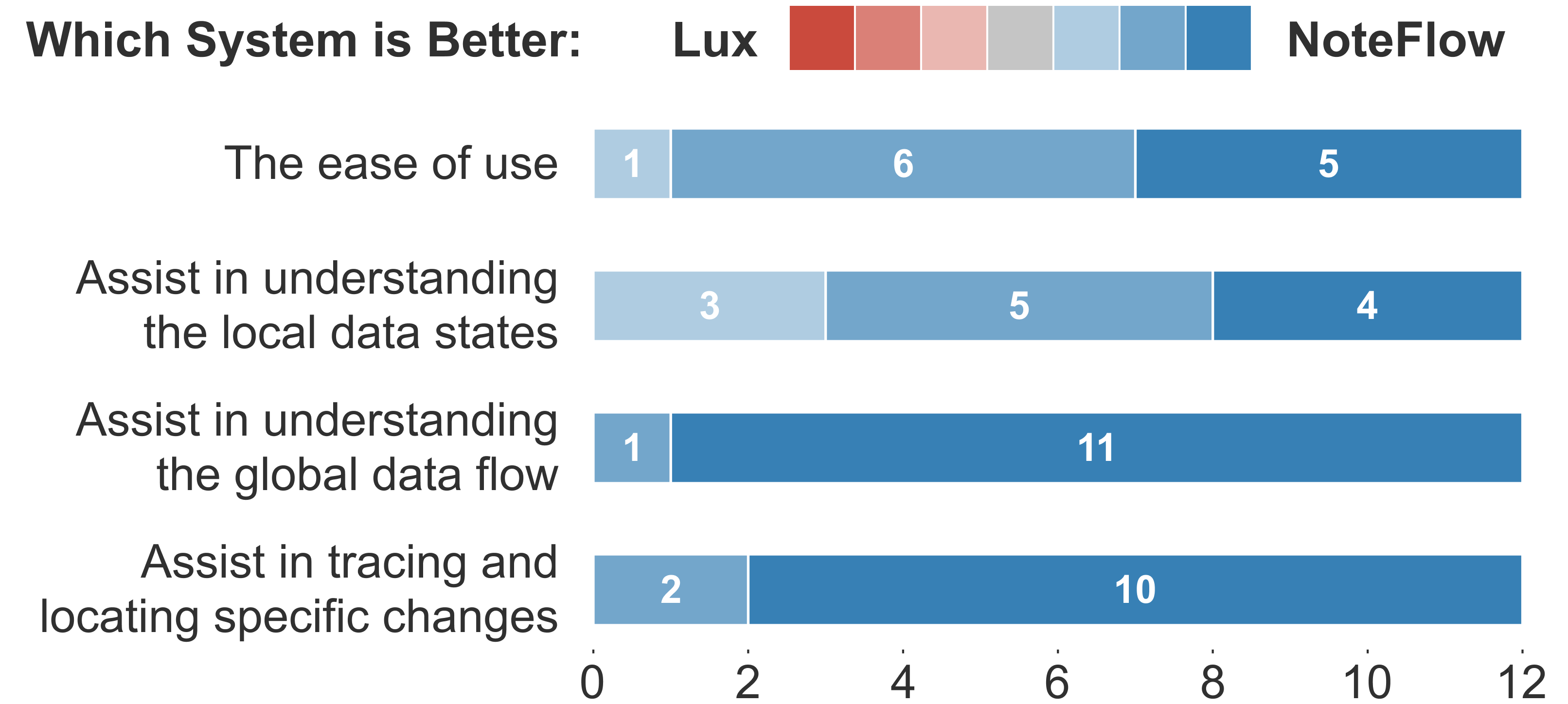}
    \caption{The result of the post-study questionnaire. }
    \Description{The figure summarizes responses from a post-study questionnaire comparing Lux and NoteFlow. It uses horizontal stacked bars on a scale from ``Lux is better'' to ``NoteFlow is better'' for four aspects: ease of use, understanding local data states, understanding the global data flow, and tracing/locating specific changes. Across all questions, responses predominantly favor NoteFlow, with the strongest preference for helping users understand the global data flow and for tracing and locating specific changes.}
    \label{fig:question_result}
\end{figure}

\textbf{Post-study questionnaires and interview (about 30 minutes): }
After completing both sessions, participants filled out a seven-point Likert scale questionnaire, rating their experiences with \system{} and Lux (\autoref{fig:question_result}). 
A post-study interview was conducted to explore the reasons behind their ratings and gather additional feedback on using both systems for data flow tracing.
In total, the experiment lasted approximately 75 minutes. 
Each participant received compensation of \$15 for their time.

\subsection{Task Performance Analysis}
\begin{figure}[!htb]
    \centering
    \includegraphics[width=\linewidth]{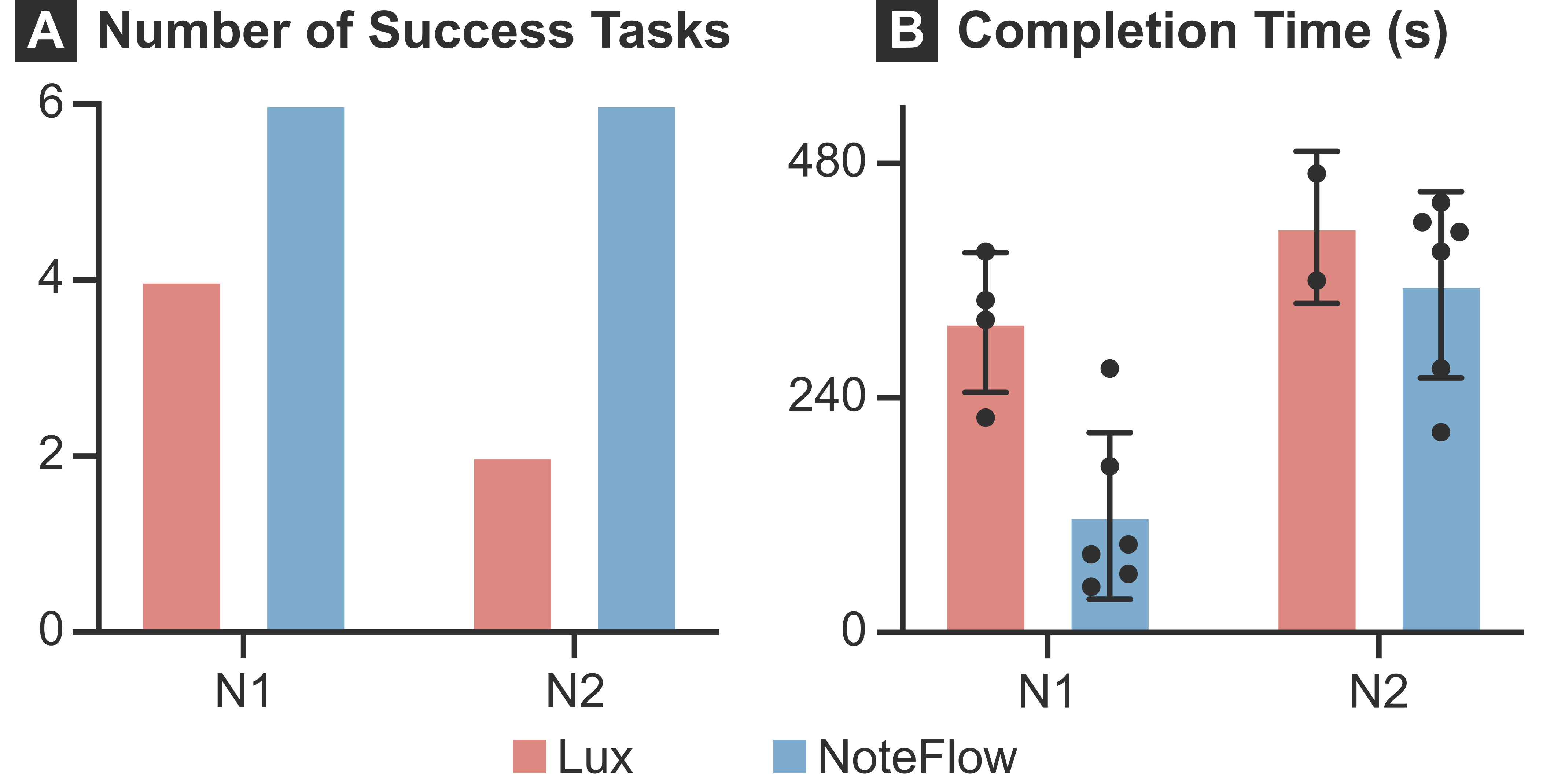}
    \caption{The number of successful tasks and the time used to complete these tasks among the success cases with Lux and \system{}.}
    \Description{
    The figure compares task performance between Lux and NoteFlow across two notebooks (N1 and N2). Panel A is a bar chart showing the number of successfully completed tasks:  NoteFlow achieves 6 successes in both N1 and N2, while Lux achieves 4 in N1 and 2 in N2. Panel B shows completion time (seconds) for the successful cases using bars with individual participant points and error bars; across both notebooks, NoteFlow completes tasks faster on average than Lux, and N2 generally takes longer than N1 for both tools.
    }
    \label{fig:case_result}
\end{figure}

As shown in~\autoref{fig:case_result}, all participants successfully identified the anomaly and located the responsible line of code using \system{}, regardless of whether they were working with N1 or N2. 
In contrast, participants encountered varying levels of difficulty when using Lux: only 4 out of 6 succeeded with N1 and just 2 out of 6 with N2. 
Among those who succeeded with Lux, task completion times were generally longer compared to \system{}. 

To better understand these differences, we analyzed participants' actions during the task. 
In both systems, participants often chose not to inspect data states immediately after each execution, especially when the operation appeared reasonable. 
This reflects a common EDA pattern, where users check data at key stages rather than at every step. 
Most participants tried to revisit the data flow to identify the anomaly's origin when it emerged at the end, which is the key step resulting in the differences in success rates and task durations.

Lux required more manual effort to revisit the data flow from a particular attribute.
When participants reviewed the recommendation history, they often found that the default recommendations did not cover the anomaly attribute, as the code did not reference it. 
As a result, participants frequently needed to specify custom chart intents and rerun cells to recover overwritten data. 
On average, participants inserted code 6.1 times to activate recommendations and specify intents, and reran the entire notebook 2.8 times. 

In contrast, \system{} streamlined the tracing process. 
Participants could select a chart relevant to the anomaly attribute and trace it to follow the data flow backward. 
Charts were expanded only at steps where the underlying dataframe changed, making it easier to locate the responsible transformation.
Based on our observations, participants used the tracing function 2.5 times and inspected recommended charts of cells 3.0 times on average, and none of the participants needed to rerun the cells when using \system{}. 

\subsection{Post-Study Feedback}
\label{sec:feedback}

The post-study questionnaire results are shown in~\autoref{fig:question_result}. 
We further summarize participant feedback below.

\textbf{\system{} offers a low-barrier interface. }
Most participants (11/12) agreed that \system{} is far better or better than Lux in terms of ease of use. 
Six participants noted that both systems had relatively low learning curves: \system{} requires some interface familiarization, and Lux involves learning APIs for recommendations. 
However, they highlighted that \system{}'s interactions were easier to remember. 
P2, P6, and P12 described \system{} as ``intuitive and natural,'' making it more memorable, whereas Lux often required referring to documentation. 
Additionally, all participants mentioned that Lux's code-based interaction was more demanding than \system{}'s graphical interface.
As P3 explained, ``I need to think about which attributes to view, how to display them, and how to use the API.'' 
In contrast, \system{} allowed for quick exploration: ``I just click and instantly get a lot of information, then look for what's meaningful.''

\textbf{\system{}'s broad chart coverage with interactive filtering supports comprehensive local data state awareness.}
All participants agreed that both \system{} and Lux were useful for understanding the local data states.  
Most participants (9/12) found \system{} to be far better or better, primarily due to its recommendation scope and interaction style.
Lux provides smart but limited chart recommendations based on user intent and data context, which are effective for discovering key insights but limited for general state inspection. 
As P11 noted, ``Sometimes the charts are not aligned with the attribute I want to see, and I have to write code to specify what I want.''
In contrast, \system{} surfaces a wide range of basic charts across attributes and combinations, allowing users to quickly explore by filtering them.
P8 remarked that ``viewing many charts is necessary to understand the data,'' and \system{} made this easier. 
P6 further highlighted the difference in interaction: in \system{}, she could ``simply click the filtering panel to find the desired attributes,'' while Lux required ``manual scanning of charts and writing intents.''

\textbf{\system{}'s consistent and continuous view facilitates understanding global data flow.}
Most participants (11/12) found \system{} far more effective than Lux for understanding global data flow, and appreciated the idea of using consistent chart encodings to monitor how dataframes evolved. 
P2 described this as ``highly intuitive,'' noting that charts can be ``easily spotted at a glance'' and the consistent encoding facilitated ``comparisons across intermediate dataframes,'' supporting a clear understanding of the overall flow.
The node-link graph, combined with auto-collapsed charts, was also praised for presenting a continuous, high-level overview. 
P11 noted that ``The data evolving process is entirely shown in the diagram, and only the changed charts expand, so I can instantly see how the data changes and where it shifts.''
P4 further commented, ``The different colors for nodes and links make the data flow clear.''
In contrast, Lux lacked a unified global view, requiring users to remember data states across cells. 
P2 and P10 found it difficult: ``It is easy to lose track of earlier states, and I have to scroll back and forth to review them.'' 
Participants also noted Lux's inconsistency across cells, with P4 mentioning that ``the charts recommended in different cells are distinct, and it is hard to follow without a clear link between them.'' 
P5 described the process as ``disjointed,'' adding that ``Lux lacks the continuity to display data flow effectively.''

\textbf{\system{} enables efficient locating of specific data changes through flexible tracing chart selection.}
Most participants (10/12) found \system{} more effective than Lux for locating specific data changes. 
P1 noted that with \system{}, he could ``easily see where the anomalies occurred and link the charts to the relevant code line,'' which minimized the need to inspect each line individually.
Participants also appreciated that \system{} allowed tracing from any chart at any stage without re-executing cells.
P7 emphasized this flexibility: ``I don't need to rerun the code to recover intermediate data, and the tracing functionality is especially suitable for following a selected data perspective at any point.''
In contrast, Lux required more manual effort. 
Participants often needed to split code cells, rerun earlier steps, or write custom code to access intermediate results. 
P3 remarked, ``I had to run each cell separately and split them, which took many iterations to find the issue.''
P7 described the process as a ``binary search,'' involving repeated trial and error to locate the source of anomalies.

\textbf{Global awareness supports better understanding of local states.}
Several participants (P4, P11, P12) highlighted that \system{}'s visualization of the overall data flow also improved their understanding of local data states. 
P4 explained, ``When the relationships between dataframes are complex, to be clear about the local state, I also need to know where I am in the data flow, including the source of the current dataframe and how it has changed. \system{} clearly highlights the evolving process and provides a consistent view to compare with previous states.''
In notebooks with multiple evolving dataframes and non-linear dependencies, \system{} helped users stay oriented—answering the question, ``Where are we?'' in the data flow. 

\textbf{Tracing forward and backward for dynamic monitoring.}
Our initial use case for \system{} envisioned users executing scripts after completing a coding stage and, upon encountering anomalies, using \system{} to trace backward and identify the source of these issues. 
This approach aligns with the feedback discussed above, which emphasizes the system's effectiveness in pinpointing specific data changes.
An unexpected insight emerged from six participants during the user study: they expressed a desire to use \system{} not only to trace backward but also to trace forward.
During EDA, they would add specific charts to the tracing list before starting transformations. 
As they proceeded with transformations, they could observe real-time updates, reflecting a dynamic, continuous data profiling approach. 
This feedback suggests that \system{} can support ongoing data profiling, aligning with concepts from previous work on continuous data profiling~\cite{autoprofiler}.

\textbf{Improve \system{} by integrating tabular views.}
At the time of our study, \system{} did not support directly displaying dataframes.
Four participants praised Lux's ability to switch between Pandas table view and chart view, and suggested that \system{} would benefit from an integrated table view to complement the existing visualizations. 
While visualizations highlight key patterns, they are by nature aggregated and may not capture all details. 
While less intuitive, data tables offer a complete and accurate representation of the data. 
P6 and P7 noted that data tables offered a fuller picture of the data state and were helpful for validating current data conditions.
We have since integrated this feature into the current system.

\section{Field Study} %

While the controlled study demonstrates the effectiveness of \system{}, it restricts participants' actions to predefined tasks, which does not fully capture how analysts behave in natural EDA workflows. To address this gap, we conducted a one-month field study in real-world settings.

\subsection{Study Setup}
The field study was designed to understand how \system{} would be used in authentic EDA workflows beyond the controlled study setting, focusing on real-world analytical scenarios.

\textbf{Participants.} The study involved two professional data scientists, referred to as \textbf{DA} and \textbf{DB}, who both regularly encounter new datasets and conduct EDA as an integral part of their work. 

\begin{itemize}[leftmargin=*]
    \item \textbf{DA} is a sports analyst with over five years of experience analyzing football and basketball data. His workflow involves receiving new datasets approximately every two months and conducting EDA sessions about once every two weeks to process the data and extract insights that inform sports training.
    \item \textbf{DB} is a data analyst specializing in telecommunication networks, with about three years of experience. He typically receives new datasets on a monthly basis, each containing large-scale user measurement report data, and performs EDA to explore and prepare the data for downstream machine learning tasks.
\end{itemize}
 
\textbf{Procedure.} We asked both participants to deploy and use \system{} in their real workflows over the course of one month. To monitor progress and ensure smooth usage, we conducted weekly interviews to check whether they were still using the system, identify any engineering issues, and provide timely fixes when necessary. At the end of the month, we carried out a final in-depth interview focusing on (1) how they typically conduct EDA in real scenarios, (2) how they used \system{} compared to working without it, and (3) other qualitative feedback on the system design choices.

Based on the field study observations and interviews, we summarize our findings in the following subsections.

\subsection{EDA Workflow and Use of \system{}}

This subsection focuses on participants' EDA workflows and how \system{} supported their activities in practice.

\subsubsection{Participants Follow a Three-Stage EDA Workflow.} Both DA and DB described their EDA process as three recurring stages:
\begin{itemize}[leftmargin=*]
\item \textbf{Data reading and profiling}: loading datasets, inspecting metadata, and checking dataframes by examining attributes, value distributions, missing values, and sample records.
\item \textbf{Data transformation}: cleaning and restructuring the data to address issues such as missing values or inconsistent formats, and reshaping it into formats suitable for later analysis. 
\item \textbf{Data analysis}: applying operations such as grouping or filtering and visualizing results to obtain insights and identify patterns. 
\end{itemize}
This sequence is not strictly linear. 
DA noted that analysis often raised new questions, requiring additional profiling and transformations to generate new views of the data. 
DB emphasized that his workflow ultimately aimed to prepare the data for machine learning, which required a final transformation step of reshaping data into training-ready formats.
We next discuss how \system{} supported these workflows compared to working without it.

\subsubsection{Local Chart Recommendations Reduce the Overhead of Inspection. }
\system{} provides a local inspection feature after each cell execution, allowing users to immediately examine the resulting data.
Both DA and DB reported that they have used this feature across all three stages of EDA, though the types of recommended charts varied by stage.
During data loading, they browsed the visualization and table panels to quickly inspect column types, value distributions, and missing data. During transformations, both found it useful that charts of transformed attributes were prioritized, allowing them to immediately verify processed results. 
During analysis, they found that the recommended charts appeared immediately after relevant operations. 
For example, DA could view group-by results right after a group-by operation, while DB obtained scatter plots after selecting subsets and correlation heatmaps after running correlation analysis.

In summary, these features reduced the overhead of calling functions such as \texttt{info()} or \texttt{describe()} or writing visualization code manually. 
DB noted that he often forgets matplotlib parameters and appreciated not having to search documentation or prompt an LLM to generate plots.
DA further noted that without \system{}, he relied on custom print statements or manual visualization code, which were time-consuming and often led him to skip checks and miss potential issues. 
With the integrated table and chart panels, he was more willing to inspect additional columns, reducing the risk of overlooking unexpected problems.

\subsubsection{Chart Tracing Builds a Mental Model of Data Evolution.}
In addition to local inspection, \system{} provides a tracing feature that allows users to select a recommended chart of any executed cell to trace its global evolution. 
DA and DB reported using this feature primarily in three scenarios: data transformations, debugging, and code reuse/understanding. 

For transformations, tracing helped them verify how specific attributes changed after processing. 
DA explained that when applying transformation operations, he often needed to confirm whether the results were correct. 
With \system{}, he could simply trace the charts with operated attributes, immediately seeing their states before and after operations and how they evolved across the notebook. 
Without the tool, he relied on manual print statements and ad-hoc rules to infer correctness, which was both time-consuming and error-prone.
For debugging, DB highlighted that chart tracing, which automatically expands changed data states and highlights the data flow path, was especially valuable for locating bugs.
He noted that a single click could reveal when problematic data first appeared, greatly reducing the manual effort of backtracking through earlier cells.
For code reuse/understanding, both scientists pointed out that they often work with new datasets that share similar but slightly different structures, which leads them to reuse past scripts. 
This process carries a memory cost, as they need to recall what the scripts originally did to the data. 
Tracing helped them quickly re-establish an understanding of how transformations affected the dataset. 
DB further noted that tracing was similarly helpful for understanding code generated by LLMs.

Summarizing across scenarios, DB noted: ``Tracing is helpful when I lack a clear mental model of the transformations applied. It gives me visibility into what happened and where it happened.''

\subsection{Feedback on Re-execution during EDA}
\label{subsection:field_study_rerun}

A key motivation of \system{} is to help analysts better manage non-linear execution patterns that arise when cells are revised and re-run in exploratory analysis. 
From the field study, we specifically examined how participants approached re-execution and how our design choices supported this process. 
We summarize their feedback in two aspects: layout preferences when visualizing re-runs, and how \system{} influenced the extent of re-execution in practice.

\textbf{Use linear layouts for re-running. }
Both scientists occasionally encountered non-linear execution when revising and re-running cells. 
We explored two design options with them (\autoref{fig:re-run}): 
(1) a stepped layout where new executions create additional attributes shown side-by-side with earlier results, 
and (2) a linear layout where re-executed cells are placed normally at the bottom, preserving a top-down flow. 
Both preferred the linear design, as the stepped layout was difficult to scale with repeated re-runs and consumed too much screen space.

\textbf{\system{} reduces re-running cells. }
DA further remarked that using \system{} reduced the extent of re-execution in his EDA workflow. He explained that he usually re-executed earlier cells to restore past data states when checking issues, which created non-linear execution. With \system{}, this behavior was reduced in two ways:
(1) easier data inspection encouraged him to check states more frequently, allowing him to catch issues earlier and avoid back-tracing; 
and (2) chart tracing helped him quickly locate problematic transformations, so he could directly modify the relevant code and re-run it once, rather than repeatedly re-running multiple cells.

\subsection{Reflection and Gaps}
\label{subsection:field_study_limit}
While overall feedback was positive, the field study uncovered several cases that exposed boundaries in \system{}'s current support.

\textbf{Chart coverage.} 
\system{} supports a set of basic charts, including histograms for single-attribute distributions, scatter plots for two-attribute correlations, line charts for temporal trends, and bar charts for group-by aggregations (e.g., x: category, y: aggregated value).
Both DA and DB agreed that the provided charts covered the vast majority of their scenarios, but noted a few infrequent cases that were not supported.
DA sometimes required multi-level group-by with multiple attributes and aggregations (e.g., grouping by both team and season and then computing mean and sum across performance metrics). 
DB mentioned occasional needs for SHAP plots or geo-heatmaps. 
Since \system{} does not currently support these cases, they had to create such visualizations manually during their EDA sessions.

\textbf{Saved intermediate data, loops, and function calls.}
One limitation observed during the field study concerned how \system{} saves intermediate data. By default, the system stores the output of every code statement that produces a dataframe, which can take up substantial space when many statements are executed. This issue became especially apparent with loops. DB sometimes processed data by iterating over groups, and in earlier versions \system{} saved a dataframe and generated a node for each iteration, resulting in excessive intermediate data and a cluttered trace view. To address this, we later modified the behavior to keep and display only the first and last iterations by default. DB further suggested that in the future, it would be useful to allow users to selectively preserve additional iterations or delete saved states that are no longer needed.
Another case noted in the study involved function calls. DB occasionally processed data using self-defined functions, but \system{} currently does not support tracing inside a function body. The system only captures data states before and after a function's execution, without visibility into the transformations that occur within the function.

Overall, these cases highlight practical boundaries of the current implementation. While they did not prevent participants from completing their tasks, they illustrate opportunities for improving scalability and flexibility in future versions of \system{}.

\section{Discussion}
In this section, we discuss the lessons learned, limitations, and future work of \system{}. 

\subsection{Lessons Learned}
Through the iterative development of \system{}, we derived two key design lessons.
First, understanding global data flow requires both identifying relationships between dataframes and effectively comparing them.
While a node-link diagram can reveal dependencies, meaningful comparison depends on \textbf{consistent visual layouts}, following the principle of comparative visualization~\cite{comparative2021}.
In an earlier version, we embedded recommended charts from different cells into a unified node-link diagram. 
Each cell offered different charts, leading to a heterogeneous node-link diagram with various charts of distinct attributes, making it difficult to interpret the overall data flow.
Users found this interface overwhelming and perceived the system primarily as a local chart recommendation tool. 
The intended global perspective was diminished due to the lack of contextual coherence across charts.
Therefore, we separated the recommendations from the node-link diagram. 
The diagram now adopts a consistent chart encoding to trace how a specific data perspective evolves across cells, enabling clearer global comparison.

Second, we learned that an \textbf{intuitive overview} of the data early in the analysis is more important than deep insights. 
Initially, we ranked recommended charts based on the importance of insights, such as strong correlations. 
These charts often dominated the top of the list throughout the session.
However, users reported confusion in the early stages of EDA, as they were still unfamiliar with the dataset and found correlation charts too abstract at this point. 
Therefore, we adjusted the ranking strategy to prioritize charts of operated attributes and their distributions. 
Insight-driven charts remain accessible in the recommendation list and can be explored via the filtering panel when users are ready to investigate more complex patterns.

\subsection{Limitations and Future Work}
We identify the main limitations of \system{} and outline directions for future improvement.

\textbf{Chart recommendation.}
\system{} recommends charts to help data analysts inspect intermediate data states.
Specifically, it recommends charts by considering observed data operations and characteristics, trying to align with the user's analytical intent. 
To surface a broad range of potentially relevant views,
\system{} incorporates several mechanisms, including providing basic charts for all attributes, prioritizing charts associated with recently operated attributes, and supporting direct table previews.
Nevertheless, certain cases can still be missed.
From a coverage perspective, some transformations may require more specialized visualizations to reveal their effects (e.g., operations that alter distributional assumptions relevant to statistical tests).
From a ranking perspective, changes to indirectly affected attributes may not appear early in the recommended list.
More fundamentally, users' analytical intent is implicit and may extend beyond what can be inferred directly from data transformations and attribute-level data characteristics, so the system's inference of user intent is inherently imperfect. 

Potential extensions include expanding the set of recommended chart types when a transformation implies specialized analytical needs, enhancing code analysis to better detect indirect effects, and incorporating interactive mechanisms that allow users to refine or customize the charts (e.g., through lightweight configuration interfaces or LLM-assisted chart generation).

\textbf{Visualizing subtle data changes.}
NoteFlow visualizes the traced charts as data evolves, helping analysts understand the global progression of dataframe states and the effects of data transformations. 
However, some transformations may produce minimal or imperceptible visual differences, even though they meaningfully alter statistical properties or downstream analytical validity. 
For example, rounding operations or small numeric adjustments may not visibly shift a plot, despite altering distributional characteristics. 
Visualizations alone cannot always reveal these subtle effects.
While \system{} provides direct table previews, identifying changes across tables often requires careful manual comparison, and the preview shows only the first few rows, which may not surface localized modifications. 
A promising direction is to incorporate explicit diff-based highlighting for both tables and charts, and to provide focus-guided navigation that automatically moves to or enlarges modified regions in either view.

\textbf{Implementation limitations. }
Beyond the recommendation- and visualization-related limitations, \system{} also has several implementation constraints that affect its generality and scalability. 
First, \system{} currently only supports Pandas dataframes and relies on rule-based parsing of Pandas operations. 
This poses two issues: 
(1) the system cannot handle other data formats or libraries (e.g., NumPy arrays and PyTorch tensors, which may appear in later stages of EDA and modeling workflows); 
and (2) the rule set may require frequent adaptation if Pandas updates its APIs. 
A promising direction is to extend parsing to multiple libraries and reduce reliance on rule-based approaches. 
Since tracing only requires identifying input–output relationships, operation types, and affected data attributes, future work could leverage LLMs to infer these dependencies in a more flexible and generalizable manner. 

Second, \system{} incurs storage overhead because it saves all intermediate dataframe states by default. 
This issue was also noted in the field study (\autoref{subsection:field_study_limit}), especially when loops or iterative transformations generated many states. 
We currently mitigate this by keeping only the states of the first and last iterations. 
A potential improvement is to allow users to specify which states to retain (e.g., selecting particular iterations in a loop or retaining only the most recent n dataframes) and to provide state management features such as pruning or deleting unneeded states.

Third, \system{} currently captures only the input and output states of user-defined functions, without tracing the transformations that occur inside the function body. 
A potential improvement is to provide optional function-level tracing, allowing users to selectively drill into a function to view its internal operations. 

\section{Conclusion}
In this paper, we propose \system{}, a notebook extension that leverages charts to trace data flow during exploratory data analysis.
The core idea is to use charts as intuitive perspectives of dataframes, helping users understand how data evolves throughout the analysis.
We implemented \system{} and evaluated it through both a comparative user study and a one-month field study.
The comparative study demonstrated \system{}'s effectiveness for data flow comprehension, while the field study confirmed its utility in real-world workflows.
We further discussed key insights and lessons learned from both studies, as well as the current limitations of \system{} and directions for future improvement.

\begin{acks}
This work was supported by NSFC (62421003, U22A2032, 62402428) and Ningbo Yongjiang Talent Programme (2023A-396-G). The authors also gratefully acknowledge the support of Zhejiang University Education Foundation Qizhen Scholar Foundation.
\end{acks}

\bibliographystyle{ACM-Reference-Format}
\bibliography{ref}

\end{document}